\documentclass[12pt]{iopjournal}

\usepackage{graphicx}%
\usepackage{multirow}%
\usepackage{amsmath,amssymb,amsfonts}%
\usepackage{amsthm}%
\usepackage{mathrsfs}%
\usepackage[title]{appendix}%
\usepackage{xcolor}%
\usepackage{textcomp}%
\usepackage{manyfoot}%
\usepackage{booktabs}%
\usepackage{algorithm}%
\usepackage{algorithmicx}%
\usepackage{algpseudocode}%
\usepackage{listings}%
\usepackage{quantikz}
\usepackage{subcaption}
\usepackage{adjustbox}
\usepackage{comment}
\usepackage{booktabs}
\usepackage{diagbox}
\usepackage{makecell}
\usepackage[numbers,sort&compress]{natbib}

\renewcommand{\headrulewidth}{0pt}
\renewcommand{\footrulewidth}{0pt}

\renewcommand{\articletype}[1]{}

\begin{document}

\articletype{Paper} %	 e.g. Paper, Letter, Topical Review...

\title{Deployment of Large IQP Circuit Born Machines on Qubit-Limited Hardware Using Noise Injection}

\author{Ju-Young Ryu\orcid{0009-0007-7293-9403}, Wooyeong Song\orcid{0000-0002-7127-5170}, Kwangil Bae\orcid{0009-0007-7001-671X}, Wonhyuk Lee\orcid{0000-0002-1571-9638} and IlKwon~Sohn$^{*}$\orcid{0000-0001-5321-2427}}

\affil{Quantum Network Research Center, Korea Institute of Science and Technology Information, Daejeon, Republic of Korea}

\affil{$^*$Author to whom any correspondence should be addressed.}

\email{d2estiny@kisti.re.kr}

\keywords{Instantaneous Quantum Polynomial Circuit, Noise-induced Simulability, Quantum Circuit Born Machines}

\begin{abstract}
Quantum circuit Born machines (QCBMs) built with instantaneous quantum polynomial (IQP) circuits have loss functions that can be efficiently estimated classically, but may require quantum computers during inference. Thus, classical computers can train large IQP QCBMs that require more qubits than available quantum hardware provides. This motivates research into how qubit-limited quantum computers can be used to sample from larger IQP QCBMs. Leveraging insights from classical simulation of noisy IQP circuits, we propose a deployment protocol for IQP QCBMs that injects noise at the logical level to probabilistically decompose the original circuit into smaller clusters that can be sampled on qubit-limited quantum processors. We also apply lightweight resampling of known noisy qubits to mitigate the impact of the introduced error. Numerical simulations of shallow IQP QCBMs and IBM Heron experiments deploying a 316-qubit IQP circuit onto 156-qubit hardware show that our protocol, combined with error mitigation, reproduces selected graph-informed correlators more accurately than a classical baseline with accurate single-qubit marginals. In the hardware experiment, mitigation also reduces the maximum mean discrepancy from the target distribution.
\end{abstract}

\section{Introduction}\label{sec1}
Instantaneous quantum polynomial (IQP) circuit-based quantum circuit Born machines (QCBMs) are a promising candidate for near-term quantum generative modeling. They can be trained classically using a specific loss function and data-dependent initialization, while inference is delegated to quantum hardware \cite{recio2025train}. An efficient implementation allowing GPU acceleration and backpropagation for gradient calculation has been released \cite{armengol2025iqpopt}, making the model accessible. They have been proven to be universal when a subset of qubits are traced over \cite{kurkin2025universality}, showing their expressivity. Recent experiments on noisy intermediate-scale quantum hardware deployed shallow IQP QCBMs successfully and preserved properties of the generated data, showing their feasibility on near-term hardware \cite{ballo2026shallow}. Since the loss functions needed to train these models can be calculated on classical computers, models larger than available hardware can be trained. Therefore, inference of large IQP QCBMs when there is only access to smaller quantum computers is a major bottleneck. 

A natural solution to this bottleneck is to partition a large circuit into smaller fragments to be executed using multiple quantum processors - the goal of distributed quantum computation (DQC). However, existing DQC methods are not ideal for the deployment of classically trained large IQP QCBMs on near-term hardware. There are two representative methods for distributing a quantum computation task. Teleportation-based distributed quantum computing requires entanglement pairs to be distributed between separate nodes and classical communication during computation \cite{bennett1993teleporting, cirac1999distributed, gottesman1999demonstrating, eisert2000optimal}. Technology that implements this is not yet readily available. On the other hand, quasi-probability decomposition-based distributed quantum computing realizes operations between separate nodes by decomposing them into a mixture of local operations \cite{peng2020simulating, mitarai2021constructing, lowe2023fast, schmitt2025cutting}. When applied to weak simulation of quantum circuits, this method comes at the expense of higher shot costs related to the $l_1$-norm of the quasi-probability decomposition \cite{onishi2026weak}.

We propose a noise injection-based deployment protocol in order to provide an alternative path towards the deployment of IQP QCBMs. This protocol approximates the noiseless IQP Born distribution as a set of nontrivial sampling tasks that can be performed on smaller quantum hardware. This is achieved by injecting logical error into the original circuit, utilizing ideas from classical simulation \cite{rajakumar2025polynomial}. However, the injected noise degrades sample quality and changes correlations between bits. We design lightweight error mitigation methods aligned with the percolation phenomenon in noisy IQP circuits to alleviate the effects of the injected noise. These methods operate by resampling specific bits that are known to be dominantly affected by noise.

We evaluated our protocol on IQP QCBMs trained for graph generation \cite{ballo2026shallow} using noiseless classical simulations, and on a 316-qubit IQP sampling task deployed on a 156-qubit IBM Heron processor. We used the product of the target distribution’s single-bit marginals as a classical baseline to assess the recovery of correlations beyond independent bit sampling. In both simulations and hardware experiments, the mitigated protocol achieved lower mean absolute errors (MAE) for selected graph-informed Pauli Z word expectation values than this baseline. In the hardware experiment, LocalCondEM reduced these errors by approximately 46\% for edge pairs and 38\% for connected triples. More broadly, this work illustrates how noise-induced simulability can be constructively repurposed as a deployment tool for qubit-limited quantum hardware.

\section{Preliminaries}\label{sec2}

We first discuss the IQP QCBM and the classical simulation algorithm that provides the main technical impetus behind our proposed method.

\subsection{Quantum circuit Born machines and instantaneous quantum polynomial circuits}

Quantum circuit Born machines (QCBMs) are generative models that, given a training dataset of bit-strings sampled i.i.d. from an underlying distribution, learn a quantum circuit which prepares a quantum state whose Born distribution is close to the underlying distribution~\cite{liu2018differentiable, benedetti2019generative}. By preparing a quantum state using the trained QCBM and measuring it, one could generate a new sample from the data distribution. Deploying a QCBM is equivalent to sampling from the Born distribution of the quantum state prepared by the trained quantum circuit.

\begin{figure}
    \centering
    \begin{quantikz}
        \lstick{$|+\rangle$} & \gate[3]{D} & \gate{H} & \meter{}\\
        \lstick{$\vdots$} & \setwiretype{n}& \vdots & \vdots \\
        \lstick{$|+\rangle$} &  & \gate{H} & \meter{}
    \end{quantikz}
    \caption{An $n$-qubit instantaneous quantum polynomial (IQP) circuit consists of an initial state $|+\rangle^{\otimes n}$, a unitary $D$ diagonal in the $Z$ basis, and measurement in the $X$ basis.}
    \label{fig:iqp_circuit}
\end{figure}
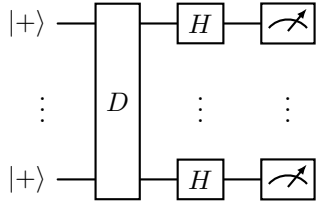

An $n$-qubit instantaneous quantum polynomial (IQP) circuit is a circuit consisting of: (1) initial state $|+\rangle ^{\otimes n}$, (2) gates $D$ diagonal in the $Z$ basis, and (3) measurement in the $X$ basis~\cite{shepherd2009temporally} as shown in Figure~\ref{fig:iqp_circuit}. These circuits are non-universal in that arbitrary unitaries cannot be realized using IQP circuits to a chosen precision. Despite this non-universality, sampling classically from the Born distributions of IQP circuits has been proven to be difficult under complexity-theoretic assumptions~\cite{bremner2011classical, bremner2016average, fujii2017commuting}. 

Due to the classical hardness of IQP sampling, creating a QCBM with parameterized IQP circuits has been proposed as a promising candidate for a quantum generative model. Interestingly, a classically efficient method for calculating a loss function for training such models has been suggested~\cite{recio2025train}. This method utilizes the fact that expectation values of Pauli Z words can be efficiently estimated for an IQP circuit~\cite{nest2009simulating}. An estimator of the square of the maximum mean discrepancy (MMD)~\cite{gretton2012kernel} between the underlying data distribution and the IQP circuit's Born distribution can be written as a linear combination of Pauli Z word expectations~\cite{rudolph2024trainability}, thus allowing classical estimation of a loss function for QCBM training. In later sections, we use these methods to estimate Pauli Z expectation values and MMDs to compare sampling methods.

\subsection{Percolation-based classical simulation of noisy IQP circuit sampling}

The IQP QCBM deployment protocol proposed by this study is based on a classical algorithm for simulating noisy IQP circuit sampling~\cite{rajakumar2025polynomial}. This classical algorithm simulates sampling from the Born distribution of an IQP circuit where single-qubit Pauli noise channels act on each qubit after every diagonal unitary layer. Define a single-qubit Pauli noise channel $N_{p_x, p_y, p_z}$ as follows:
\begin{equation}
    N_{p_x,p_y,p_z}(\rho) = (1-p_x-p_y-p_z)(\rho) + p_xX(\rho)X+p_yY(\rho)Y+p_zZ(\rho)Z
\end{equation} 
Each single-qubit noise channel is decomposed into a classical mixture of two components:
\begin{equation}
    N_{p_x, p_y, p_z}(\rho) = (1-2p)N_1(\rho) + 2p(N_2 \circ N_{0, 0, 1/2})(\rho)
\end{equation}
\begin{equation*}
    p = p_z+\min(p_x, p_y), N_1 = \begin{cases}
        N_{\frac{|p_x-p_y|}{1-2p}, 0, 0}, p_x \ge p_y\\
        N_{0, \frac{|p_x-p_y|}{1-2p}, 0}, \text{otherwise}
    \end{cases}, N_2 = N_{\frac{\min(p_x, p_y)}{p}, 0, 0}
\end{equation*}
The operator $\circ$ denotes the sequential application of two channels. $N_1$ is either a bit-flip channel or a bit-phase-flip channel depending on the $p_x$ and $p_y$ values, while $N_2 \circ N_{0, 0, 1/2}$ is a bit-flip channel applied after a completely dephasing channel in the computational basis. The noise channels can be simulated by choosing an error pattern for each measurement sample, where $N_1$ or $N_2 \circ N_{0, 0, 1/2}$ is applied to each noise location with probability $1-2p$ or $2p$, respectively. $N_{0, 0, 1/2}$ commutes with other Pauli channels and diagonal gates that make up the IQP circuit. Therefore, on qubits where there is at least one $N_{0, 0, 1/2}$ channel, such channels propagate to the beginning of the circuit to transform the initial state to the maximally mixed state $N_{0,0,1/2}(|+\rangle\langle +|)=I/2$.

After commuting all $N_{0, 0, 1/2}$ channels to the beginning, each affected qubit is a uniformly random computational basis state. The following diagonal gates that act on these computational basis states are equal to operations on the remainder of their support, where the sign of the rotation angle is determined by the computational basis state. For instance, $RZZ(\theta)(|1\rangle\langle1| \otimes \rho)RZZ^\dagger(\theta) = |1\rangle\langle1|\otimes RZ(-\theta)\rho RZ^\dagger(-\theta)$. Gates acting only on removed qubits contribute sample-dependent phases and can be omitted. The remaining $N_1$ and $N_2$ noise channels acting on the computational basis state can be classically simulated as bit-flip channels. The original circuit is thus divided into separable circuits because a subset of qubits does not become entangled with others. The separable circuits can be simulated on a classical computer using conventional density matrix methods when they are of a tractable size. 

The interaction between the noise and the circuit can be represented using a phenomenon called node percolation. Node percolation is the behavior of a graph in which nodes are removed from the graph with a certain probability. Consider a graph where the nodes correspond to qubits and the edges correspond to two-qubit gates in the IQP circuit. In this work, we call this graph the interaction graph $G_I = (N_I, E_I)$. $N_I$ is the set of qubits and $E_I$ is the set of two-qubit gates. The process of dividing a large $n$-qubit quantum circuit into smaller separable clusters due to certain qubits becoming maximally mixed states can be interpreted as node percolation. 
This is because for $d$ independent noise layers, each node becomes detached from the original circuit with probability $1-(1-2p)^d$, where $p=p_z+\min(p_x, p_y)$. 
The size of the largest connected component decreases as $2p$ increases, eventually splitting into smaller clusters of size $O(\log n)$. The classical simulation algorithm operates in this regime, where $O(\log n)$ size circuit fragments are efficiently simulated on a classical computer and the fragment results are concatenated to create a measurement sample from the noisy distribution.

We utilize the node percolation mechanism of the classical simulation algorithm to deploy IQP QCBMs. While the classical algorithm simulates hardware noise, we deliberately inject additional logical dephasing error into the original circuit and utilize its effect. This logical noise approximates the noiseless Born distribution and reduces the circuit size to fit the available quantum hardware.

\section{Proposed deployment protocol}\label{sec3}

In this section, we describe the proposed deployment protocol in detail. We first discuss the design of the injected error and the method to determine its strength. We also explain how the injected error is simulated to approximate the original circuit into a sampling task that can be run on available quantum hardware. We conclude this section by introducing two error mitigation methods that can be used to reduce the approximation error caused by the extra noise.

\subsection{Noise-induced percolation}

\begin{figure}
    \centering
    \includegraphics[width=0.8\textwidth]{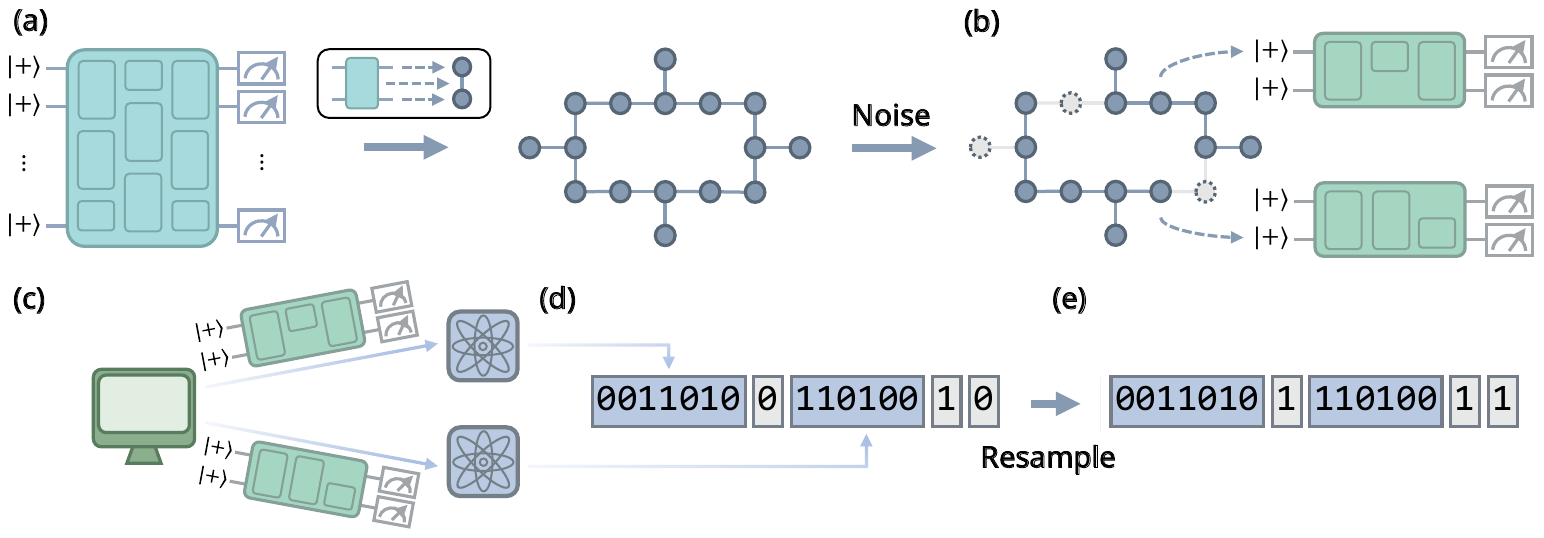}
    \caption{Overview of the proposed noise injection-based deployment of IQP QCBMs. (a) A trained IQP QCBM is represented using an interaction graph. (b) Logical noise is injected into the circuit, which can be represented as probabilistically removing nodes from the interaction graph and creating smaller IQP circuits. (c) These smaller IQP circuits are run on qubit-limited quantum computers. (d) The outputs from each computer are concatenated along with the bits corresponding to the removed nodes. (e) If error mitigation is used, the removed node bits are resampled from classically calculated marginal distributions.}
    \label{fig:protocol_summary}
\end{figure}

We propose a noise injection-based IQP QCBM deployment protocol based on the noisy IQP sampling simulation algorithm discussed earlier. The proposed protocol operates in a regime where the removal probability of each qubit from the $n$-qubit interaction graph is strong enough to divide the interaction graph into hardware-sized components, but not necessarily strong enough to reduce all components to $O(\log n)$ size for efficient classical simulation. Figure~\ref{fig:protocol_summary} summarizes the proposed deployment protocol and the subsequent error mitigation step.

The protocol begins by first determining the structure of the injected noise in the circuit. Considering that the available quantum computers are also noisy, we utilize a different decomposition of the single-qubit Pauli channel by separating the noise model of the available quantum computer $N_{p_x, p_y, p_z}$ from the completely dephasing channel $N_{0, 0, 1/2}$. We denote by $p_p$ the node removal probability over the full depth-$d$ circuit and by $p_p^l$ the per-layer logical dephasing probability, so that $p_p = 1-(1-p_p^l)^d$. The subscript $p$ refers to node percolation. If the single-qubit noise model of the simulated circuit $N(\cdot)$ is: 
\begin{equation}\label{eq:injected-noise}
    N(\rho)=(1-p^l_p)N_{p_x,p_y,p_z}(\rho) + p^l_p N_{0,0,1/2}(\rho) = N_{(1-p^l_p)p_x,(1-p^l_p)p_y,(1-p^l_p)p_z+p^l_p/2}(\rho) 
\end{equation}
each qubit becomes a maximally mixed state with probability $p_p = 1-(1-p_p^l)^d$. Relative to the hardware noise channel, the effective channel has reduced X and Y error weights and an increased Z dephasing weight. The reduction in X and Y error weights is due to the fact that the logically introduced noise only contains Z errors. Note that we sample from a deliberately noise-injected approximation to the target IQP circuit, not from the original noiseless circuit.

Next, the node removal probability is determined. We estimate the node removal probability $p_p$ for which the largest connected component of $G_I$ remains within the hardware capacity $N_q$ with probability at least $p_{succ}$. 
This is achieved by calculating the average largest connected component size over an ensemble of graphs with the same distribution of nodes degrees. This approach is the standard configuration model percolation theory~\cite{molloy1995critical, molloy1998size, newman2018networks}, which entails solving a self-consistency equation using numerical methods. Details of the percolation theory are given in Appendix~\ref{secAppendixPercTheory}. In practice, the estimated $p_p$ value can create clusters too large for the available quantum computers. To prevent this, we perform a grid search near the estimated node removal probability to fine-tune the noise strength so that the fraction of samples with giant cluster size within the capacity, or success probability, is estimated to be at least $p_{succ}$. To test node removal probability $p_p$, a subset of nodes $R$ is randomly sampled by selecting each node with probability $p_p$. Then the size of the largest connected cluster $L_{max}(G_I\setminus R)$ is counted, where nodes in $R$ are removed from the interaction graph $G_I$. This is repeated a set number of times to calculate an empirical estimate of the success probability. The smallest $p_p$ within the search grid such that the success probability is larger than $p_{succ}$ is the fine-tuning result. A choice of a $p_{succ}$ close to 1 will result in high levels of injected noise, and so this value can be used to control the tradeoff between sample quality and the classical noise pattern sampling overhead. In this work, we use $p_{succ} = 0.9$ to balance injected noise level and success probability. Note that only selecting node removal patterns that result in fragments within capacity conditions the noise pattern distribution on successful fragmentation. This results in the implemented protocol sampling from a distribution that approximates the IQP QCBM with the noise model given in Equation~\ref{eq:injected-noise}.

Finally, the effect of the injected noise is simulated and the resulting circuit fragments are executed. The noise channels are simulated by either choosing $N_{p_x,p_y,p_z}$ or $N_{0,0,1/2}$ in each noise location with probability $1-p^l_p$ or $p^l_p$ respectively. Following the classical noisy IQP sampling simulation algorithm, $N_{0,0,1/2}$ channels are propagated to the beginning and the original circuit is divided into separable circuits. After running the separated circuits on available quantum computers, measurement bit-strings from each connected cluster are collected along with the uniformly random bits from the maximally mixed states to be concatenated into a single noisy sample. Note that depending on the size of the separated circuits, multiple small circuits may be run on a single quantum computer, or they could be classically simulated. The multiple sampling tasks could also be distributed to multiple separate quantum computers to be run in parallel. This allows for a flexible mixture of classical and quantum computation. A complete summary of the algorithm is given in Appendix \ref{secAppendixAlgo}.

\subsection{Error mitigation via resampling}

Each sample produced by the protocol comes with a known partition of qubits into surviving nodes $S$ and removed nodes $R$. As discussed in the protocol description, bits from $S$, $x_S$, are obtained by executing the corresponding IQP fragments on the available hardware, and therefore inherit the physical noise of the fragment processors. However, bits from $R$, $x_R$, arise from logical dephasing events and are eventually sampled as uniformly random outcomes. We can attempt to mitigate the dominant distortion by resampling the removed bits using information from the target IQP circuit and, when available, the measured surviving bits.

Exact reconstruction conditioned on the surviving node measurement results is not expected to be efficient in general. Conditioning an IQP output distribution on a specified outcome of $S$ can be viewed as a postselection operation, and postselected IQP circuits have the power of PostBQP, equivalently PP \cite{bremner2011classical}. This provides complexity-theoretic evidence that generic exact conditional sampling from $P(x_R\mid x_S)$ is intractable, especially when the conditioning event has exponentially small probability. We therefore consider two efficient approximations of $P(x_R\mid x_S)$. ProdEM (\ref{sec:prodem}) discards $x_S$ and resamples each bit independently from its estimated ideal single-bit marginal. LocalCondEM (\ref{sec:localcondem}) conditions each removed bit on its surviving neighbors, recovering local correlations at a cost that grows with the node degree.

\subsubsection{ProdEM}\label{sec:prodem}

A simple method for mitigating errors in single-bit marginal distributions is to resample the removed bits using the marginal distributions inferred from each qubit's Pauli Z expectation value evaluated using the final output bit-strings, 
\begin{equation}
    P_\text{ProdEM}(x_R) = \prod_{j\in R} P_j(x_j), P_j(0) = \frac{1+\langle Z_j\rangle_{ideal}}{2}, P_j(1) = \frac{1-\langle Z_j \rangle_{ideal}}{2}
\end{equation}
Using the software \texttt{Iqpopt} \cite{armengol2025iqpopt} developed for the training of IQP QCBMs on classical computers, one can estimate the single-qubit Pauli Z expectation values of an $n$-qubit IQP circuit to error $\epsilon = \text{poly}(n^{-1})$ using $\text{poly}(n)$ time and space. Thus, this method has the benefit of not using extra quantum resources. It does not explicitly model correlations involving removed bits, although correlations entirely within surviving clusters may still be preserved by the fragment samples. We denote this method ProdEM in the following sections.

There are some performance guarantees of ProdEM. Consider a scenario where the single-bit marginal distributions are known without error and the hardware that is used to deploy the IQP QCBM has no quantum noise. Then ProdEM is guaranteed to match or reduce certain weight-2 Pauli Z word expectation value errors compared to the unmitigated distribution. More specifically, the error of Pauli Z words on interaction graph edges that are not a part of a triangle is multiplied by a factor of $\sin^2\theta_{ij}$ for circuits consisting of RZ and RZZ gates, where $\theta_{ij}$ is the rotation angle of the RZZ gate acting on the edge $(i,j)$. Under the same assumptions, ProdEM also reduces each pair’s expectation value error relative to the product baseline by a factor \(1-\alpha_{ij}\), where \(\alpha_{ij}\) is the probability that both endpoints survive. Details on the derivation are given in Appendix~\ref{secAppendixEMAnalysis}.

\subsubsection{LocalCondEM}\label{sec:localcondem}

We can also use information from the surviving neighbors of a removed node to construct the marginal distribution. For a removed node $i\in R$, let $B_i$ be the set of its neighbors that were not removed. Since the measurement outcomes $x_{B_i}$ are known, we approximate the removed bit distribution by the local conditional marginal $P(x_i\mid x_{B_i})$ of the ideal IQP Born distribution. If $B_i=\emptyset$, this reduces to the single-qubit marginal used in ProdEM.

The conditional marginal can be computed from Pauli Z word expectation values.
Using
\begin{equation}
    |b\rangle\langle b|=\prod_{j\in B_i}\frac{I+(-1)^{b_j}Z_j}{2},
\end{equation}
the conditional \(Z_i\) expectation is
\begin{equation}
\langle Z_i \rangle_{x_{B_i}} = \frac{\langle Z_i \otimes |x_{B_i}\rangle \langle x_{B_i}|\rangle}{\langle|x_{B_i}\rangle \langle x_{B_i}|\rangle} = \frac{\sum_{W\subseteq B_i}\left(\prod_{j\in W}(-1)^{x_j}\right)\langle Z_i Z_W\rangle_{\mathrm{ideal}}
}{
\sum_{W\subseteq B_i}\left(\prod_{j\in W}(-1)^{x_j}\right)\langle Z_W\rangle_{\mathrm{ideal}}
}.
\end{equation}
The ideal expectation values are estimated using \texttt{Iqpopt}. We then resample $x_i$ from the distribution
\begin{equation}
P(x_i=b\mid x_{B_i})
=
\frac{1+(-1)^b\langle Z_i\rangle_{x_{B_i}}}{2}, b\in \{0, 1\}.
\end{equation}
We denote this method LocalCondEM.

For a fixed removed node $i$ with degree $\deg(i)$, this procedure requires expectation values for all Pauli words supported on subsets of $B_i$, both with and without $Z_i$. Thus, the number of required expectation values is at most $2^{|B_i|+1}\le 2^{\deg(i)+1}$. Averaging over measurement outcomes of $B_i$ and the Monte Carlo estimation of each expectation value, the expected total variation error is at most $\epsilon$ when $O(2^{|B_i|}/\epsilon^2)$ Monte Carlo samples are used per expectation value. Across the required expectation values, the total number of Monte Carlo estimator evaluations required is $O(4^{|B_i|}/\epsilon^2)$.
Therefore, LocalCondEM is efficient for bounded-degree interaction graphs, and remains polynomial for graphs of logarithmic maximum degree. The required expectation values depend only on the local conditioning set \(B_i\), not on the observed bit values \(x_{B_i}\), so they can be cached and reused across samples. The resource bound is derived in Appendix~\ref{secAppendixLocalCondEM}.

If the conditional marginal distributions are calculated without error and the quantum hardware that is used to deploy the IQP QCBM is not affected by quantum noise, LocalCondEM also has some performance guarantees. Under these assumptions, LocalCondEM fully recovers some Pauli Z word expectation values for certain qubit removal patterns. For instance, reconstruction occurs when only one node is removed from an edge pair or when only the middle node is removed from a three-node path. More details on this property are given in Appendix~\ref{secAppendixEMAnalysis}. 

Combining the noise injection-based protocol with resampling-based error mitigation allows approximate sampling from IQP circuit Born machines with more qubits than the available hardware. In the following section, we use numerical simulations and hardware experiments to evaluate agreement with the ideal output distribution and the accuracy of selected local Pauli Z expectation values.

\section{Performance evaluation}\label{sec4}

We evaluate the proposed protocol and the error mitigation methods introduced above using numerical simulations as well as experiments on IBM Heron hardware. In particular, we measure how accurate local correlations are through the mean absolute error (MAE) of Pauli Z word expectation values and how dissimilar the sampled distribution is to the ideal distribution using the squared maximum mean discrepancy (MMD$^2$). We show that, with the help of error mitigation, the proposed protocol has a comparable MMD$^2$ and more accurate graph-informed correlators compared to a classical baseline.

\subsection{Numerical simulation: 28-qubit QCBMs for graph generation}\label{sec4_1}

To test how well our protocol preserves local correlations, we apply the proposed protocol to shallow IQP QCBMs for generating undirected and unweighted graphs trained in a previous study \cite{ballo2026shallow}. This study trained each model using a dataset that consisted of 8-node graphs drawn from different ensembles (Bipartite/Erd\H{o}s-R{\'e}nyi) and connectivity levels (Dense/Medium/Sparse). Each IQP circuit was approximately sampled from using the noise injection-based deployment protocol, where the available quantum computers are simulated to be noiseless with qubit counts of 15, 18, and 21. This choice of model size and removal of noise in the simulated quantum computer keeps the problem size in a simulable region while isolating the impact of the artificially introduced noise.

The empirical distribution of the proposed deployment protocol samples is compared to the trained model's Born distribution and a classical baseline method. The classical baseline used in the following analysis is the product of single-bit marginal distributions. The marginal distributions can be calculated through the single-qubit Pauli Z expectation values, as utilized in ProdEM. This baseline will be called `product' in the following sections. The product baseline is intentionally strong for one-body and product-like statistics because it uses ideal single-bit marginals. Improvements over the baseline indicate recovery of correlations beyond independent marginal matching.

\begin{figure}[!htb]
    \centering
    \includegraphics[width=0.9\linewidth]{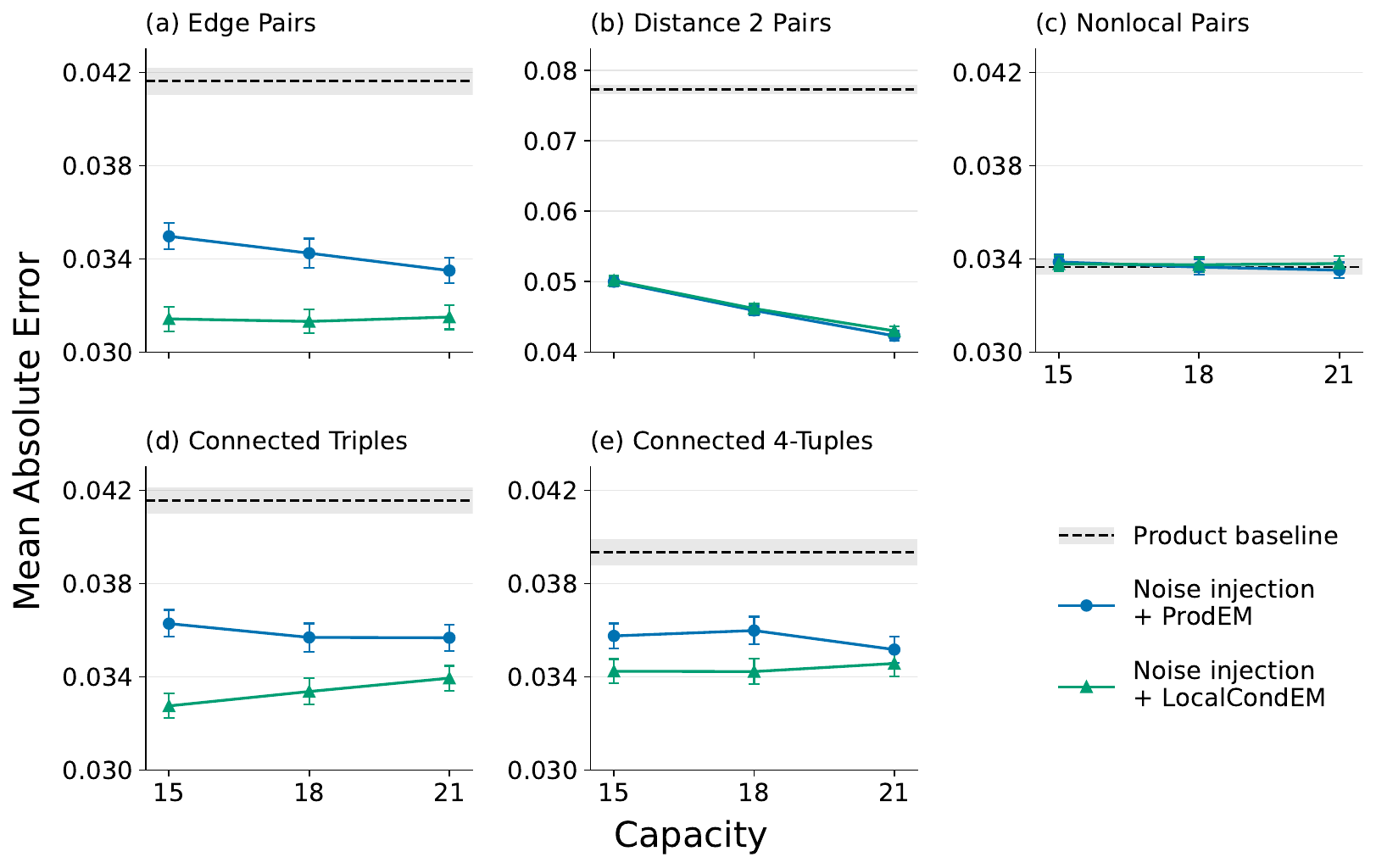}
    \caption{Graph-informed Pauli Z word expectation value errors. Markers and the dashed lines are calculated by averaging over six circuits and 50 repetitions per circuit. Error bars and the product band show 95\% percentile bootstrap intervals from 5,000 resamples stratified by circuit. Note (b) uses a different vertical scale. The protocol with error mitigation yields lower mean errors compared to the baseline on most considered local graph-structured observable families. (a) For edge pairs, LocalCondEM recovers ideal marginals whenever at most one endpoint is removed, providing an advantage over ProdEM. (b) For distance-2 pairs, the single-removal exactness guarantee does not generally apply; the two methods exhibit similar errors in these simulations. (c) Uncorrelated nonlocal pairs make the product baseline competitive. LocalCondEM shows a slight advantage over ProdEM on connected Pauli Z words of weight (d) 3 and (e) 4.}
    \label{fig:graph_correlators}
\end{figure}

We calculate Pauli Z word expectation values to measure correlation reconstruction. On a sparse interaction graph, uniformly random low-weight Pauli supports are often separated by graph distance large enough that their relevant neighborhoods weakly overlap or do not overlap. These observables are therefore frequently close to product-like statistics and may understate differences between a marginal baseline and a sampler that preserves local correlations. Therefore, we evaluate graph-informed Pauli word ensembles. Figure \ref{fig:graph_correlators} compares Pauli Z word expectation value MAEs for the two mitigated deployment methods and the product baseline. Both error mitigation methods reduce the error, becoming more accurate than the product baseline over most considered Pauli word families. Panels (a)–(c) compare weight-two observables at increasing separation. LocalCondEM provides its clearest advantage for (a) observables with supports on the interaction graph edges (`Edge Pairs'), whose ideal marginals it recovers whenever at most one endpoint is removed. This guarantee does not extend to (b) observables with supports that are distance-2 on the interaction graph (`Distance 2 Pairs'), for which the two mitigation methods exhibit similar errors. For (c) uncorrelated nonlocal observables (`Nonlocal Pairs'), the product distribution with exact single-bit marginals reproduces the ideal pair marginal, leaving no intrinsic bias to correct. The nonzero error of the baseline in panel (c) occurs from finite sampling and variance from estimating the ideal expectation values. Panels (d)–(e) examine whether the benefits extend to higher-weight observables on interaction graph connected supports. `Connected Triples' refers to weight-3 observables and `Connected 4-Tuples' refers to weight-4 observables. LocalCondEM retains an advantage in these results, although exact recovery depends on both the removal pattern and the support topology. Additional results on random Pauli Z words are included in Appendix~\ref{secAppendixRandomPaulis}, and empirical data on how these local correlations translate to domain specific sample quality metrics are included in Appendix~\ref{secAppendixSampleQuality}.

\subsection{Hardware experiment: approximate 316-qubit IQP sampling on IBM Heron}

The proposed deployment protocol makes strong assumptions about the structure of the noise within the quantum computer. To test the practical performance of the protocol for realistic quantum hardware, we apply the proposed protocol to deploy a 316-qubit IQP sampling task on 156-qubit \texttt{ibm\_marrakesh}. Note that this is a deployment and mitigation demonstration rather than a full generative model quality benchmark. The target large IQP circuit is a 316-qubit IQP circuit with RZ gates on all qubits followed by RZZ gates on the edges of the heavy-hex lattice. The interaction graph is equivalent to two 156-qubit heavy-hex lattice graphs connected by 4 intermediate nodes, which has bounded node degree. The rotation angles are sampled from $U[0, 2\pi]$ for RZ and $U[0, \pi/2]$ for RZZ in order to utilize fractional gates~\cite{ibmfractional}. The diagonal portion of the circuit begins with a layer of RZ gates, followed by layers of RZZ gates scheduled by edge coloring the interaction graph.

Before applying the protocol, one needs to model the quantum hardware's noise as identical single-qubit Pauli channels acting layer-wise on each qubit. We achieve this by running IQP circuits on small disjoint sets of the total available physical qubits, which form a partition. The $\text{MMD}^2$ values between the classically simulated distribution and samples from the quantum hardware are averaged across different random IQP circuits and partitions. This average value is minimized by changing the simulated noise model's parameters. For this IQP workload, the single-parameter depolarizing model achieved training and validation MMD$^2$ values comparable to those of other considered two- and three-parameter models, motivating its use as an effective noise model. More details on the modeling experiment can be found in Appendix~\ref{secAppendixIBMDetails}.

\begin{figure}[h]
    \centering
    \includegraphics[width=0.9\linewidth]{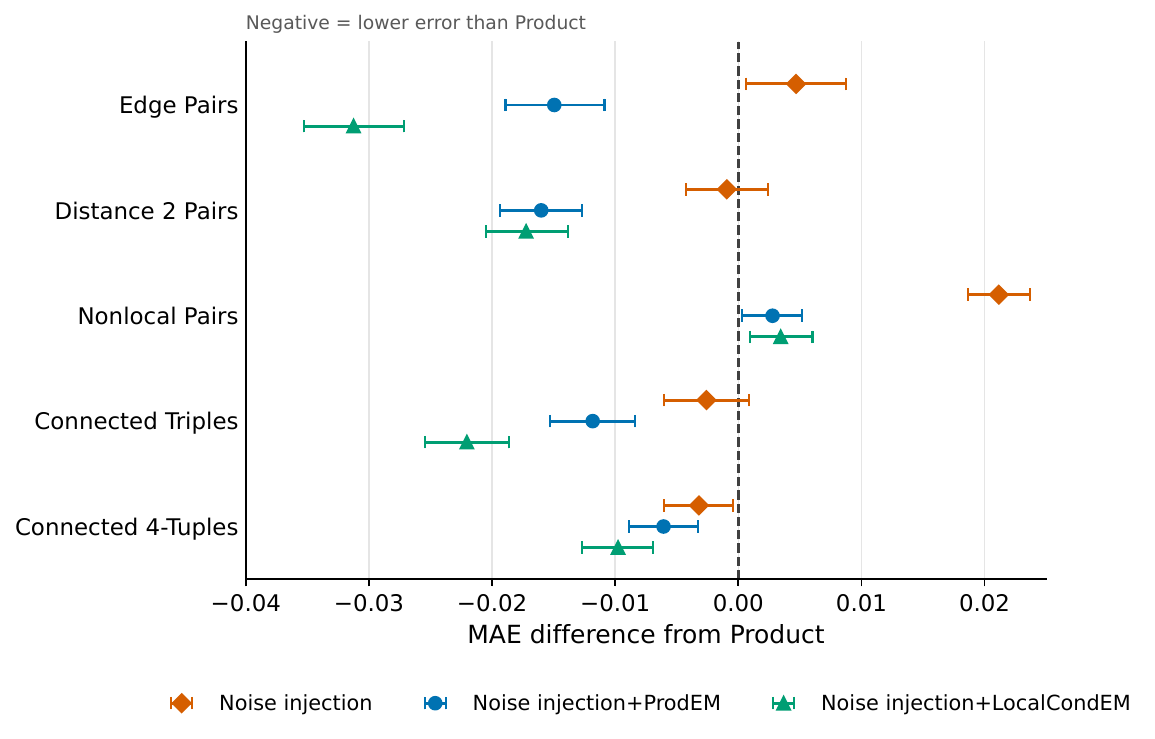}
    \caption{IBM experiment Pauli Z word expectation value error differences relative to the product baseline, where negative values indicate lower error than the baseline. Error bars indicate the recentered 95\% bootstrap percentile intervals from 5,000 independent bit-string resamples of each method and the baseline. Edge pairs have support on the interaction graph edges, distance 2 pairs have supports with distance 2 over the interaction graph, nonlocal pairs have supports with distance 4 or higher over the graph, and connected triples(4-tuples) are weight-3(4) words with support connected by the graph. Both mitigation methods yield lower errors than the product baseline for all considered families except nonlocal pairs. LocalCondEM provides its largest additional improvements over ProdEM for edge pairs and connected triples, while the product baseline has the lowest error for nonlocal pairs. These qualitative trends agree with the noiseless simulation results.}
    \label{fig:ibm_correlators}
\end{figure}

The difference between the Pauli Z word expectation value MAEs of the product baseline and each sampling method is calculated as in Figure~\ref{fig:ibm_correlators}. We report observed point estimates with recentered 95\% bootstrap confidence intervals. Negative values indicate lower error than the baseline. Both mitigation methods outperform the product baseline for edge pairs, distance 2 pairs, and connected triples and 4-tuples, whereas the unmitigated protocol performs comparably to the baseline for some of these families. The product baseline achieves the lowest error for nonlocal pairs, whose ideal connected correlations vanish. The relative performance of the mitigation methods follows the same qualitative trends as in the noiseless simulations. LocalCondEM provides its largest additional improvements over ProdEM for edge pairs and connected triples, while the differences are smaller for distance 2 pairs and connected 4-tuples. Considering that the MAE of the product baseline on edge pairs is $6.739\times 10^{-2}$ and connected triples is $5.749\times 10^{-2}$, LocalCondEM's improvement corresponds to approximately 46\% and 38\% reduction of error, respectively. The trends shown in the simulation results persist in the IBM hardware experiment despite hardware noise and the use of a simplified noise model.

Low-weight Pauli Z expectation values alone cannot indicate how well the sampled distribution agrees with the ideal distribution. To compare sampling methods at a distribution level, we calculate $\text{MMD}^2$ via \texttt{Iqpopt}, as shown in Table \ref{tab:ibm_mmd}. For each sampling method, we generated 500 bootstrap datasets by resampling bit-strings with replacement, maintaining the original dataset size. For each bootstrap dataset, we averaged 5 independent $\text{MMD}^2$ evaluations, and the 2.5th and 97.5th percentiles of these bootstrap averages are reported. Because the IBM experiment produces one sample set per method, these intervals do not capture hardware drift or calibration variability. While the protocol has a markedly high $\text{MMD}^2$ compared to other methods, both error mitigation methods reduce $\text{MMD}^2$ to values comparable to those of the product baseline.

\begin{table}
\caption{IBM experiment $\text{MMD}^2$. Values are bootstrap means and the intervals are obtained from 500 bit-string resampled datasets. The protocol without error mitigation has $\text{MMD}^2$ higher than the baseline, but error mitigation reduces it to values comparable to the baseline.}\label{tab:ibm_mmd}
\begin{tabular*}{\textwidth}{@{\extracolsep{\fill}}lcc}
\toprule Method & $\text{MMD}^2$ & [2.5th, 97.5th] Percentiles\\ 
\midrule
Product Baseline & $3.38\times 10^{-4}$ & $[3.08\times 10^{-4}, 3.72\times 10^{-4}]$ \\
Noise injection & $4.78 \times 10^{-4}$ & $[4.45\times 10^{-4}, 5.13\times 10^{-4}]$\\
Noise injection + ProdEM & $3.51\times 10^{-4}$ & $[3.21\times 10^{-4}, 3.83\times 10^{-4}]$\\
Noise injection + LocalCondEM & $3.51\times 10^{-4}$ & $[3.19\times 10^{-4}, 3.82\times 10^{-4}]$\\
\bottomrule 
\end{tabular*} 
\end{table}

Together, the simulation and hardware results show that the protocol with error mitigation reproduces selected graph-informed Pauli Z word expectation values more accurately than a product baseline with accurate single-bit marginals. In the hardware experiment, error mitigation also reduces MMD$^2$ to values comparable to those of the product baseline. These findings support the practical feasibility of combining noise injection-based deployment with resampling-based error mitigation on quantum hardware.

\section{Conclusion}\label{sec13}

This study proposes a deployment protocol for large, classically trained IQP QCBMs on qubit-limited hardware by injecting noise at the logical level and approximating the noiseless distribution. We also introduce two lightweight error mitigation methods, ProdEM and LocalCondEM, which can be used to alleviate the impact of the injected noise without extra quantum resources. The practical viability of the noise injection-based deployment protocol is demonstrated using numerical simulations and hardware experiments. ProdEM and LocalCondEM reduced MMD$^2$ evaluated relative to the target noiseless distribution in the hardware experiment to values comparable to those achieved by the product baseline. The deployment protocol with error mitigation also reproduced selected interaction graph-informed correlators more accurately than the baseline. In the hardware experiment, LocalCondEM reduced errors by approximately 46\% for edge pairs and 38\% for connected triples relative to the baseline's error.

The proposed noise injection-based deployment protocol is primarily limited by its dependence on the circuit structure and quantum error mitigation method. The strength of the injected noise to create the same success probability would likely increase when the interaction graph is robust to node removal, potentially increasing approximation error. Also, the error mitigation methods studied in this paper perform well, but have analytical guarantees that assume idealized hardware and focus on certain observables. Designing more advanced quantum error mitigation methods that are lightweight and have strong performance guarantees is another important future research direction. Furthermore, even when the noiseless target distribution is classically hard to sample from, the protocol does not guarantee that the approximate distribution retains this hardness. Identifying regimes in which hardness is preserved remains a direction for future work.

Overall, this study shows that while noise-induced simulability is often viewed as a roadblock for quantum advantage, the same mechanisms can also be repurposed to deploy quantum generative models on near-term devices. 

\funding{This research was supported by the Korea Institute of Science and Technology Information (KISTI)(K26L1M3C5)}
% This section is a list of funder names and grant numbers

\roles{Ju-Young Ryu: Conceptualization, Formal Analysis, Investigation, Methodology, Software, Visualization, Writing - original draft, Writing - review \& editing.\\
\noindent Wooyeong Song: Investigation, Methodology, Writing - review \& editing.\\
\noindent Kwangil Bae: Investigation, Methodology, Writing - review \& editing.\\
\noindent Wonhyuk Lee: Project Administration, Supervision, Writing - review \& editing.\\
\noindent Ilkwon Sohn: Investigation, Methodology, Project Administration, Supervision, Writing - review \& editing.}
% List author names and the contributions made to the article, using terms from the NISO Contributor Roles Taxonomy (CRediT) https://credit.niso.org

\data{The data and code supporting the findings of this study are publicly available at \url{https://github.com/jyryu98/noise-injection-iqp-deploy}.}
% For more information on IOP Publishing's research data policy see: https://publishingsupport.iopscience.iop.org/questions/research-data/

\begin{appendices}

\section{Giant cluster fraction estimation}\label{secAppendixPercTheory}

Node percolation theory using the configuration model is summarized here, following Newman \cite{newman2018networks}. A configuration model is the ensemble of graphs defined by a sequence of degrees $\{k_i\}$. An alternative definition is to specify the probability that a node will have degree $k$, $p_k$. When defined using probabilities, the ensemble is created by sampling each node's degree independently from the given distribution. Given the sequence of degrees, a graph is generated by creating $k_i$ stubs, or half-edges, for each node $i$ and randomly connecting the stubs. This results in self-loops and multi-edges, but their concentration decreases as the size of the graph increases.

Node percolation results in a giant connected cluster. To estimate the size of the giant cluster $L_{max}$, we first describe the probability that when following an edge, the node at the end of the edge has degree $k$. This is $kp_k/\langle k\rangle$ in the configuration model. Thus, the probability of having $k$ excess degrees in the same situation is:
\begin{equation}
    q_k = (k+1)p_{k+1}/\langle k \rangle
\end{equation}
Let $u$ be the probability that a node isn't connected to the giant cluster through a certain neighbor. Given a node is occupied, the probability of a node not being in a giant cluster is 
\begin{equation}
    \sum_k p_k u^k = g_0(u)
\end{equation}
Then the fraction of nodes in the giant cluster $S=L_{max}/n$ is 
\begin{equation}
S = \phi[1-g_0(u)]    
\end{equation}
, where $\phi$ is the probability that a node is occupied. The probability $u$ can also be described using a self-consistency equation. 
\begin{equation}\label{eq:self-consistency}
    u = \sum_{k=0}^\infty q_k(1-\phi+\phi u^k) = 1 - \phi + \phi g_1(u)
\end{equation}
This is because the node isn't connected via a neighbor to the giant cluster if the neighbor is unoccupied or the neighbor is occupied but not connected to the giant cluster. The probability that this neighbor has $k$ excess degrees is $q_k$.

In our proposed protocol, we need to estimate a value of the node removal probability $p_p = 1-\phi$ that creates a given giant cluster fraction $S$. We solve $u = 1-S\frac{1-g_1(u)}{1-g_0(u)}$ using numerical methods, then plug the $u$ value into $\phi = S/(1-g_0(u))$. More specifically, we utilize the \texttt{fsolve} function from the Python library \texttt{SciPy} \cite{2020SciPy-NMeth}, which implements an altered version of the Powell hybrid method \cite{powell1970hybrid}.

\begin{figure}[h]
    \centering
    \includegraphics[width=0.40\textwidth]{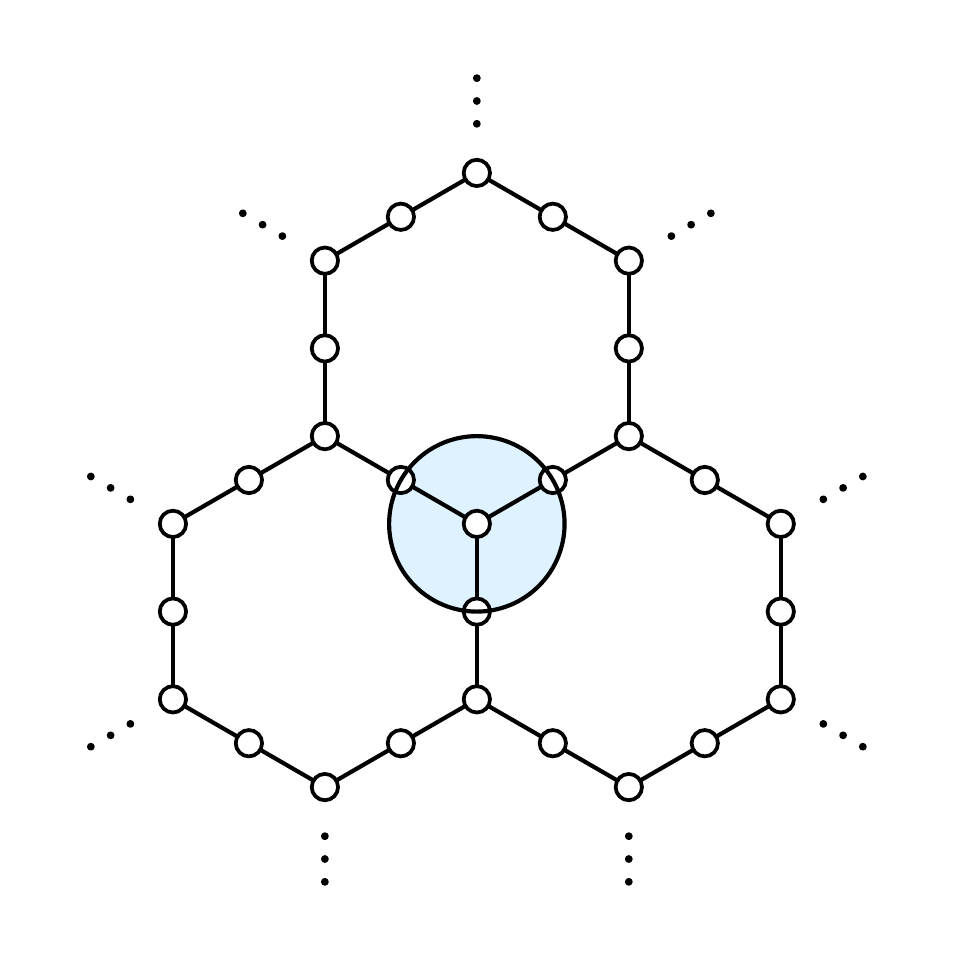}
    \caption{The heavy-hex lattice and its unit cell, indicated with the blue circle. There is one whole node with degree 3, and three half nodes with degree 2 in this unit cell.}
    \label{fig:heavy_hex}
\end{figure}

This standard configuration model can be applied to the heavy-hex lattice, the qubit connectivity structure of IBM Heron processors. By disregarding the boundary effects, one can calculate an analytical configuration model estimate of the giant cluster fraction rather than rely on numerical solvers. The unit cell of this lattice contains 1 node with degree 3 and 1.5 nodes with degree 2, as illustrated in Figure \ref{fig:heavy_hex}. Therefore, by entering the degree distribution into the consistency equation:

\begin{equation}
    u = 1-S\frac{1-\frac{1}{2}(u+u^2)}{1-\frac{1}{5}(2u^3+3u^2)}
\end{equation}
As $u\ne 1$, the consistency equation simplifies to
\begin{equation}\label{eq:heavy_hex_percolation_consistency}
    4u^3+6u^2+5Su+10S-10=0
\end{equation}
Using the Cardano formula for the cubic equation reveals one real solution
\begin{equation}\label{eq:usol}
    u(S)=-\frac{1}{2}+\sqrt[3]{\frac{18-15S}{16}+\sqrt{\Delta}} + \sqrt[3]{\frac{18-15S}{16}-\sqrt{\Delta}}
\end{equation}

\begin{equation}
    \Delta = \bigg(\frac{15S-18}{16}\bigg)^2+\bigg(\frac{5S-3}{12}\bigg)^3
\end{equation}

This solution can be used to estimate the node removal probability $p_p =1-S/(1-g_0(u(S))) = u(S)/(2+u(S))$ for a given target giant cluster fraction $S$. Interestingly, the percolation threshold is estimated to occur at $p_p^{th}= 1-\langle k\rangle/(\langle k^2\rangle -\langle k\rangle)$ $= 1/3 = u/(2+u)$, or $u=1$. Since $u$ is the probability that a node is not connected to the giant cluster through a certain neighbor, $u(S) = 1$ only in the limiting case $S = 0$. Thus, within the configuration model, the largest cluster constraint characterizes the deployment better than a sharp transition into clusters sized logarithmically small with respect to the total number of nodes.

\section{Noise injection-based deployment protocol details}\label{secAppendixAlgo}

\begin{algorithm}[H]
\caption*{\textbf{Algorithm} Noise Injection-based IQP QCBM Deployment}\label{algo_appendix}
\textbf{Input}: $n$-qubit $d$-diagonal layer IQP circuit $C$, Capacity $N_q$, Success rate $p_{\rm succ}$, Number of shots $N_{shot}$, Single-qubit noise model $N_{p_x, p_y, p_z}$\\
\textbf{Output}: Noise-injected samples and removed node sets $NS = \{(x^{(a)},R^{(a)})\}_{a=1}^{N_{\rm shot}}$
\begin{algorithmic}[1]
\State Initialize $NS = \emptyset$, make interaction graph $G_I = (N_I, E_I)$
    \State Count degrees and solve self-consistency equations for an approximate $p_p$
    \State Fine-tune $p_p$ using grid search to obtain a fraction of samples within capacity $> p_{\rm succ}$
    \While{$|NS| < N_{shot}$}
    \State Sample noise pattern $np\in \{0, 1\}^{n \times d}$, $np_{ij} \sim \text{Bernoulli}(p_p^l), p_p^l = 1-(1-p_p)^{1/d}$
    \State Remove nodes $R = \{i|\sum_j np_{ij} > 0\}$ from a copy of $G_I$
    \State Initialize $A_g = 1$ for every diagonal gate $g$ in $C$
    \If {Largest connected cluster has number of qubits larger than $N_q$}
        \State Discard $np$ and try again from line \texttt{5}
    \Else 
        \State Sample random bits $rb_i \sim U(\{0, 1\})$ for each $i\in R$.
    \EndIf
    \For{$j=1$ to $d$}
        \For {All gates $g$ in the $j^{th}$ layer of $C$}
            \If{$\text{supp}(g) \subseteq R$} $A_g = 0$
            \ElsIf{$\text{supp}(g) \cap R \ne \emptyset$} $A_g = (-1)^{\sum_{x\in \text{supp}(g)\cap R}rb_x}$
            \EndIf
        \EndFor
        \For{$i=1$ to $n$}
            \If {$np_{ij}=0$ and $i \in R$} $e\sim \text{Bernoulli}(p_x+p_y), rb_i \leftarrow rb_i \oplus e$
            \EndIf
        \EndFor
    \EndFor
    \For {Each connected cluster $G_I^k = (N_I^k, E_I^k)$}
        \State Initialize circuit $C^k$ with qubits indexed by $N_I^k$ and a Hadamard layer
        \For {Each gate $g(\theta)$ in $C$}
        \If {$\text{supp} (g) \cap N_I^k \ne \emptyset$} Apply $g(\theta \times A_g)$ to $C^k$ on $\text{supp} (g) \cap N_I^k$
        \EndIf
        \EndFor
        \State Measure $C^k$ on a quantum computer in the X basis and sample one bit-string $\vec{b}^k$
    \EndFor 
    \State Resample removed node bits $rrb_i \sim U(\{0, 1\}), \forall i \in R$
    \State Concatenate $\{\vec{b}^k\}$ and $\{rrb_i\}$ to get bit string $x$, insert $(x, R)$ into $NS$ 
    \EndWhile
\end{algorithmic}
\end{algorithm}

Note that the surviving support of any gate is only a member of a single connected component, even for multi-qubit gates. Thus, each modified gate is applied to a unique fragment circuit in each sample. Determining the connected clusters in lines \texttt{3} and \texttt{8} is performed through Breadth-First Search, which has time complexity $O(|N_I|+|E_I|)$. Therefore, this process is efficient with respect to the number of qubits. 

\section{LocalCondEM resource analysis}\label{secAppendixLocalCondEM}

Consider a removed node $i$ and the set of its surviving neighbors $B_i$. LocalCondEM calculates $\langle Z_i\rangle_{B_i}$. It is estimated through \texttt{Iqpopt}, so we only have access to an estimator $\langle \hat{Z_i}\rangle_{B_i}$ defined as:
\begin{equation}
    \langle \hat{Z_i}\rangle_{B_i} = \begin{cases}
        \text{clip}(\hat{n}/\hat{d}, -1, 1), \ \hat{d} \ne 0\\
        0, \ \hat{d} = 0
    \end{cases}
\end{equation}
\begin{equation}
    \hat{n} = \sum_{W\subseteq B_i} \left(\prod_{j \in W} (-1)^{x_j}\right)\langle\widehat{Z_i Z_W}\rangle, 
    \ \hat{d} = \sum_{W\subseteq B_i} \left(\prod_{j \in W} (-1)^{x_j}\right)\langle\widehat{Z_W}\rangle
\end{equation}
$\text{clip}(a,b,c), b<c$ outputs $b$($c$) when $a$ is smaller(larger) than $b$($c$), otherwise outputs $a$. The $\langle \hat{Z} \rangle$ estimators are each calculated using $M$ independent Monte Carlo samples following the methods from Section 4.1 of Recio-Armengol, Ahmed, and Bowles~\cite{recio2025train} and may be correlated if the estimators share Monte Carlo samples.
%and are independent of each other.
$\langle \hat{Z} \rangle$ estimators, and by extension $\hat{n}, \hat{d}$ are unbiased estimators. Also, when each Pauli Z word expectation value is estimated using $M$ Monte Carlo samples, their variance is upper bounded by $1/M$. Then the total variation distance between the ideal and estimated conditional distributions is:
\begin{equation}
    d_{TV}\left(\text{Bern}\left(\frac{1+\langle Z_i\rangle_{B_i}}{2}\right), \text{Bern}\left(\frac{1+\langle \hat{Z_i}\rangle_{B_i}}{2}\right)\right) = \frac{1}{2}|\langle Z_i\rangle_{B_i} - \langle \hat{Z_i}\rangle_{B_i}|
\end{equation}
$\text{Bern}$ is the Bernoulli distribution. Define $n, d$ as the corresponding exact quantities with exact Pauli Z word expectation values. Then $n = d\langle Z_i\rangle_{B_i}, d = 2^{|B_i|}P(x_{B_i})$. For the defined $\langle \hat{Z_i}\rangle_{B_i}, \hat{n}, \hat{d}$ and their ideal counterparts $\langle Z_i\rangle_{B_i}, n, d$,
\begin{equation}
    d|\langle \hat{Z_i}\rangle_{B_i}-\langle Z_i\rangle_{B_i}| \le 2(|\hat{n} - n| + |\hat{d} - d|)
\end{equation}
When $\hat{d} > 0$, this can be seen by dividing cases by the values of $\hat{n}/\hat{d}$ and describing $d(\langle \hat{Z_i}\rangle_{B_i}-\langle Z_i\rangle_{B_i})$. When $\hat{d} \le 0$, 
\begin{equation}
    d|\langle \hat{Z_i}\rangle_{B_i}-\langle Z_i\rangle_{B_i}| \le 2d \le 2|\hat{d}-d| \le 2(|\hat{n}-n| + |\hat{d}-d|)
\end{equation}
Note that $\langle Z_i\rangle_{B_i},n,d,$ and their estimator versions are all dependent on $x_{B_i}$, but this is suppressed for readability. We are interested in the expected conditional total variation error averaged over conditioning outcomes,
\begin{equation}\label{eq:dtv-ub}
    \mathbb{E}[d_{TV}] = \frac{1}{2}\sum_{x_{B_i}}P(x_{B_i}) \mathbb{E}_{MC}[|\langle \hat{Z_i}\rangle_{B_i}-\langle Z_i\rangle_{B_i}|] \le \sum_{x_{B_i}} \frac{1}{2^{|B_i|}} \mathbb{E}_{MC}[|\hat{n}-n|+|\hat{d}-d|]
\end{equation}
$\mathbb{E}_{MC}$ denotes averaging over the Monte Carlo estimation process and $P(x_{B_i}) = d/2^{|B_i|}$ is the ideal probability to measure subsystem $B_i$ as $x_{B_i}$. Note that:
\begin{equation}
    \hat{n}-n = \sum_{W\subseteq B_i} (-1)^{\sum_{j\in W} x_j} (\langle \widehat{Z_i Z_W}\rangle - \langle Z_iZ_W\rangle)
\end{equation}
To bound the sum of mean squared errors, consider
\begin{align}
    \sum_{x_{B_i}}(\hat{n}-n)^2 & = \sum_{x_{B_i}} \sum_{W,V\subseteq B_i} (-1)^{\sum_{j\in W} x_j}(-1)^{\sum_{k\in V} x_k} (\langle \widehat{Z_i Z_W}\rangle - \langle Z_iZ_W\rangle)(\langle \widehat{Z_i Z_V}\rangle - \langle Z_iZ_V\rangle)\\
    & = 2^{|B_i|}\sum_{W \subseteq B_i} (\langle \widehat{Z_i Z_W}\rangle - \langle Z_iZ_W\rangle)^2
\end{align}
The second line follows from the orthogonality of parity functions. Since $\langle \widehat{Z_i Z_W}\rangle$ is unbiased and has variance upper bounded by $1/M$,
\begin{equation}
    \sum_{x_{B_i}}\mathbb{E}_{MC}[(\hat{n}-n)^2] \le 4^{|B_i|}/M
\end{equation}
Utilizing this to bound the first term in Equation~\ref{eq:dtv-ub},
\begin{align}
    \sum_{x_{B_i}}\mathbb{E}_{MC}[|\hat{n}-n|] & \le \sum_{x_{B_i}} \sqrt{\mathbb{E}_{MC}[|\hat{n} - n|^2]} \\
    & \le \sqrt{2^{|B_i|} \sum_{x_{B_i}} \mathbb{E}_{MC}[|\hat{n}-n|^2]} \\
    & \le 2^{|B_i|}\sqrt{2^{|B_i|}/M}
\end{align}
The same argument applies for $|\hat{d}-d|$, and so
\begin{equation}
    \mathbb{E}[d_{TV}] \le 2\sqrt{2^{|B_i|}/M}
\end{equation}

For this to be less than $\epsilon$, $O(2^{|B_i|}/\epsilon^2)$ Monte Carlo samples per Pauli Z word are sufficient. The total number of Pauli Z words is $O(2^{|B_i|})$, so the total cost is then upper bounded by $O(4^{|B_i|}/\epsilon^2)$. Thus, on bounded degree graphs, LocalCondEM is efficient. Note that this total variation error is relative to the ideal LocalCondEM, not the ideal Born distribution.

\section{Error mitigation performance analysis}\label{secAppendixEMAnalysis}

Assume a scenario where the relevant marginal distributions are known without error and the quantum computer used for deployment is unaffected by noise. Then the performance of ProdEM and LocalCondEM can be analyzed for certain local Pauli Z word expectation values. The analysis utilizes the fact that for a fixed removal pattern, dephasing channels can propagate to the end of the circuit. Tracing out the removed qubits then leaves the ideal marginal Born distribution on the surviving qubits, provided fragment execution is noiseless.

\subsection{ProdEM improvement over the product baseline}

First, we compare ProdEM and the baseline. For a pair of qubits indexed by $(i, j)$, let $\alpha_{ij}$ be the probability that both qubits survive the percolation process. For independent removal of nodes with probability $p_p$ and disregarding conditioning on node removal patterns that result in fragments within capacity, $\alpha_{ij} = (1-p_p)^2$. Then without error mitigation,

\begin{equation}
    \langle Z_i Z_j\rangle_{NoEM} = \alpha_{ij}\langle Z_iZ_j\rangle_{ideal}
\end{equation}
With ProdEM,
\begin{equation}
    \langle Z_i Z_j\rangle_{ProdEM} = (1-\alpha_{ij})\langle Z_i \rangle_{ideal} \langle Z_j \rangle_{ideal} + \alpha_{ij} \langle Z_i Z_j \rangle_{ideal}
\end{equation}
The product baseline is
\begin{equation}
    \langle Z_i Z_j \rangle_{Product} = \langle Z_i \rangle_{ideal} \langle Z_j \rangle_{ideal}
\end{equation}
Then the absolute error between the product baseline and the ideal distribution is
\begin{equation}
    \text{Err(Product)} = |\langle Z_i \rangle_{ideal} \langle Z_j \rangle_{ideal}-\langle Z_i Z_j \rangle_{ideal}|
\end{equation}
and the absolute error between ProdEM and the ideal distribution is
\begin{equation}
    \text{Err(ProdEM)} = (1-\alpha_{ij})|\langle Z_i \rangle_{ideal} \langle Z_j \rangle_{ideal}-\langle Z_i Z_j \rangle_{ideal}|
\end{equation}
Thus, ProdEM's weight-2 Pauli Z expectation values match those of the product baseline or are more accurate under some assumptions.

\subsection{ProdEM improvement over the unmitigated protocol}

Next, we compare ProdEM with the unmitigated protocol. Let $\mathcal{S}_a$ be the set of gates whose generators anticommute with $X_a$, and $\Omega$ be the set comprised of nonempty sets of indices $\omega \subseteq \mathcal{S}_a$ such that $\prod_{j \in \omega} Z_j = I$. Also let $\theta_j, j\in \mathcal{S}_a,$ be the rotation angle of the diagonal gate indexed by $j$. Then the exact expression for Pauli Z expectation values of an IQP QCBM given by Recio-Armengol, Ahmed, and Bowles~\cite{recio2025train} is: 

\begin{equation}
    \langle Z_a \rangle_{ideal} = \prod_{j \in \mathcal{S}_a} \cos(\theta_j) + \sum_{\omega \in \Omega} \prod_{j \in \mathcal{S}_a, j \notin \omega} \cos(\theta_j) \prod_{j \in \omega} i\sin(-\theta_j)
\end{equation}
Assume that the IQP circuit's diagonal unitary consists of only RZ and RZZ gates. Then the single-qubit Pauli Z expectation value is:
\begin{equation}
    \langle Z_i \rangle_{ideal} = \cos(\theta_i)\prod_{j \neq i} \cos(\theta_{ij})
\end{equation}

We assume that any qubit pairs $(i,j)\notin E_I$ without gates have zero angle, $\theta_{ij} = 0$. Consider a pair of qubits $(i, j)$ such that $i$ and $j$ are neighbors, but do not have other common neighbors. Denote the set of neighbors of node $i$ by $\Gamma(i)$. Then $\Omega = \emptyset$ and
\begin{equation}
    \langle Z_i Z_j \rangle_{ideal} = \cos(\theta_i) \cos(\theta_j) \prod_{k \in \Gamma(i)\setminus \{j\}} \cos(\theta_{ik}) \prod_{k \in \Gamma(j)\setminus \{i\}} \cos(\theta_{jk})
\end{equation}
This means that for these specific qubit pairs,
\begin{equation}
    \langle Z_i \rangle_{ideal} \langle Z_j \rangle_{ideal} = \cos^2\theta_{ij} \langle Z_i Z_j \rangle_{ideal}
\end{equation}
The ProdEM absolute error is then
\begin{equation}
    \text{Err(ProdEM)} = (1-\alpha_{ij})|\langle Z_i Z_j\rangle_{ideal}|\sin^2\theta_{ij}  = \text{Err(NoEM)}\sin^2 \theta_{ij} 
\end{equation}

Thus, ProdEM improves by a factor of $\sin^2\theta_{ij}$ on weight-2 Pauli Z word expectation values where $i$ and $j$ do not have a shared third neighbor. Note that this applies to all edge pairs on a heavy-hex lattice, since the lattice does not have triangles.

\subsection{Exact reconstruction conditions for LocalCondEM}

Lastly, we consider when LocalCondEM can exactly fix errors. Fix a removed set $R$ and surviving set $S$. For each $i\in R$, let $B_i\subseteq S$ denote its surviving neighbors. Under the assumptions stated at the beginning of this section, LocalCondEM produces
\begin{equation}
    P_{LocalCondEM, R}(x_R,x_S)=P(x_S)\prod_{i\in R}P(x_i\mid x_{B_i}),
\end{equation}
where $P$ is the ideal output distribution. Let $W$ denote the support of a Pauli Z word. If no node in $W$ is removed, then $W\subseteq S$ and the surviving marginal guarantees $P_{LocalCondEM, R}(x_W)=P(x_W)$. If exactly one node $i\in W$ is removed and $T=W\setminus\{i\}\subseteq B_i$, then
\begin{equation}
\begin{aligned}
P_{LocalCondEM, R}(x_i,x_T)
&=\sum_{x_{S\setminus T}}P(x_S)P(x_i\mid x_{B_i})\\
&=\sum_{x_{B_i\setminus T}}P(x_{B_i})P(x_i\mid x_{B_i})\\
&=P(x_i,x_T).
\end{aligned}    
\end{equation}
Thus, these removal patterns reconstruct the entire marginal on $W$, and consequently its Pauli word expectation value, exactly. Other removal patterns may introduce bias. Define the event
\begin{equation}
    \mathcal G_W= \{R\cap W=\varnothing\} \cup \bigcup_{i\in W} \{R\cap W=\{i\},W\setminus\{i\}\subseteq B_i\}.
\end{equation}
Since Pauli-word expectation values lie in $[-1,1]$, the absolute bias for any fixed mask is at most $2$. Averaging over masks therefore gives
\begin{equation}
    \left|\langle Z_W\rangle_{\mathrm{LocalCondEM}}-\langle Z_W\rangle_{ideal}\right| \leq 2\Pr(\mathcal G_W^c).
\end{equation}
For an edge pair $W=\{i,j\}$, conditioning on all surviving neighbors guarantees exact reconstruction unless both endpoints are removed. Hence,
\begin{equation}
    \left| \langle Z_iZ_j\rangle_{\mathrm{LocalCondEM}}-\langle Z_iZ_j\rangle_{ideal}\right| \leq 2\Pr(i,j\in R).
\end{equation}
For independent node removal with probability $p_p$, this bound becomes $2p_p^2$. If masks are conditioned on satisfying a capacity constraint, the joint removal probability under the accepted-mask distribution must be used instead.

\section{28-Qubit IQP QCBM simulation details}\label{secAppendix28QDetails}

Detailed resource estimates calculated using 10,000 simulated percolation patterns are shown in Table~\ref{tab:28q-resource}. All six models share the same interaction graph, so percolation is simulated over this single graph. The theoretical $p_p$ calculated using the methods detailed in Appendix~\ref{secAppendixPercTheory} is lower than the fine-tuned $p_p$, showing the need for fine-tuning. The theoretical $p_p$ values use the actual degree distribution of the finite 28-qubit heavy-hex interaction graph. Boundary nodes modify this distribution relative to the idealized infinite lattice distribution assumed in Equation~\ref{eq:usol}, leading to different estimates. The success probability is kept approximately constant at 0.9 during the fine-tuning step, giving an expected sampling overhead of roughly 1/0.9 $\approx$ 1.11 raw attempts per accepted sample. Note that determining whether or not a sample will be discarded happens during the classical preprocessing, so this overhead does not result in extra quantum resources.

\begin{table}[h]
\caption{28-qubit model percolation deployment resource estimates. The theoretical $p_p$ predicted by Equation~\ref{eq:self-consistency} using the actual degree distribution of the interaction graph is lower than the fine-tuned $p_p$, illustrating the importance of the fine-tuning step. The fine-tuned node removal probability decreases as capacity increases, showing that less noise is introduced when the available hardware is larger. The mean largest cluster is much smaller than the hardware capacity while the 95th percentile largest cluster size for valid shots within hardware capacity is close to the capacity since the success probability is kept close to 90\% by design.}\label{tab:resource_estimate}

\begin{tabular}{c|c|c|c|c|c}
\toprule Capacity & \begin{tabular}[c]{@{}c@{}}Theoretical \\ $p_p$\end{tabular} & \begin{tabular}[c]{@{}c@{}}Fine-tuned \\ $p_p$\end{tabular} & \begin{tabular}[c]{@{}c@{}}Mean \\ Largest Cluster\end{tabular} & \begin{tabular}[c]{@{}c@{}}95th Percentile \\ Largest Cluster\end{tabular} & \begin{tabular}[c]{@{}c@{}}Success \\ Probability\end{tabular}\\ 
\midrule 15 & 6.2725$\times 10^{-2}$ & 0.2767 & 8.654 & 14 & 0.8905 \\
18 & 4.581$\times 10^{-2}$ & 0.2258 & 10.5175 & 17 & 0.8994 \\
21 & 2.621$\times 10^{-2}$ & 0.1842 & 12.4780 & 20 & 0.8914 \\
\bottomrule 
\end{tabular}

\label{tab:28q-resource}
\end{table}

50 independent sets of 512 samples were created using each sampling method. The noise injection-based deployment protocol was repeated until the target sample size was met. The same samples from the protocol were used for error mitigation to isolate the effect of the post-processing. Fine-tuning was performed by changing the result of the self-consistency equation with steps of 0.001, simulating percolation 2048 times, then checking if more than 90\% of the samples were within capacity. 

Pauli Z word expectation values are used to check if the proposed protocol reconstructs correlations, which the product baseline does not in general. We average errors over different ensembles of Pauli Z words, 50 repetitions, and across the six different circuits. For an interaction graph $G_I = (N_I, E_I)$, edge pairs sample from $\{Z_e|e\in E_I\}$, distance 2 pairs consider $\{Z_{ij}|\exists k\in N_I \ s.t. \ (i,k), (j,k) \in E_I \}$, nonlocal pairs consider pairs of nodes with distance at least 4 along the graph, and connected triples(4-tuples) consider weight 3(4) words where the corresponding nodes are connected via the graph. For weight-1 or 2 correlator comparisons, the whole set of Pauli words satisfying each condition was used unless the total exceeded 500 words, in which case 500 randomly selected words were used. For graph-informed higher weight Pauli words, random connected growth was used to generate up to 500 distinct supports. We evaluate the mean absolute error (MAE) $\text{Err}(X_{method}) = \sum_{W\in \mathcal{W}} | \langle Z_W \rangle_X-\langle Z_W \rangle|/|\mathcal{W}|$ for each sampling method and Pauli Z word set $\mathcal{W}$, where the ideal value $\langle Z_W \rangle$ is estimated by \texttt{Iqpopt} using 10,000 samples and $\langle Z_W \rangle_X$ is estimated from $X_{method}$. To compare how accurately each sampling method recreates the correlation, we calculate $\Delta = \text{Err}(X_{method})-\text{Err}(X_{Product})$. These observable sets and calculated ideal expectation values are fixed, so the intervals condition on these choices and exclude uncertainty during observable selection and \texttt{Iqpopt} estimation. We generated 5,000 stratified bootstrap replicates by resampling repetitions within each circuit and averaging the resulting circuit means. The 2.5th and 97.5th percentiles define the plotted intervals.

\section{Sample quality analysis}\label{secAppendixSampleQuality}

Following the analysis from  Balló-Gimbernat et al.~\cite{ballo2026shallow}, four different properties of the samples generated from Section~\ref{sec4_1} are calculated to quantify the quality of the samples. (1) The average density $\bar{\rho}$ is the average fraction of all possible edges that are present in a generated graph. (2) The bipartite accuracy $BP\%$ is the percentage of generated graphs that are bipartite. (3) Expected bipartivity $\bar{\beta}\in[0,1]$ is a related value that is 1 when bipartite. (4) The total variation distance (TVD) between degree distributions is the dissimilarity between the degree distribution of graphs generated by the protocol and the graphs generated by the ideal circuit. Since the goal is to generate samples with properties similar to samples from the ideal distribution, the absolute error with respect to graphs sampled from the ideal distribution is calculated for metrics 1-3.

Sample quality errors were averaged with equal weight across the six fixed circuits, each evaluated over 50 repetitions of 512 samples. For the reference values from the ideal distribution, we sampled the same number of samples using \texttt{PennyLane}~\cite{bergholm2018pennylane}. We generated 5,000 stratified bootstrap replicates by resampling repetitions within each circuit and averaging the resulting circuit means. The 2.5th and 97.5th percentiles define the plotted intervals. Within each circuit, the same repetition indices were resampled for all methods, which shared the ideal reference set for each repetition. A single Product baseline was reused across all capacities.

\begin{figure}
    \centering
    \includegraphics[width=0.85\linewidth]{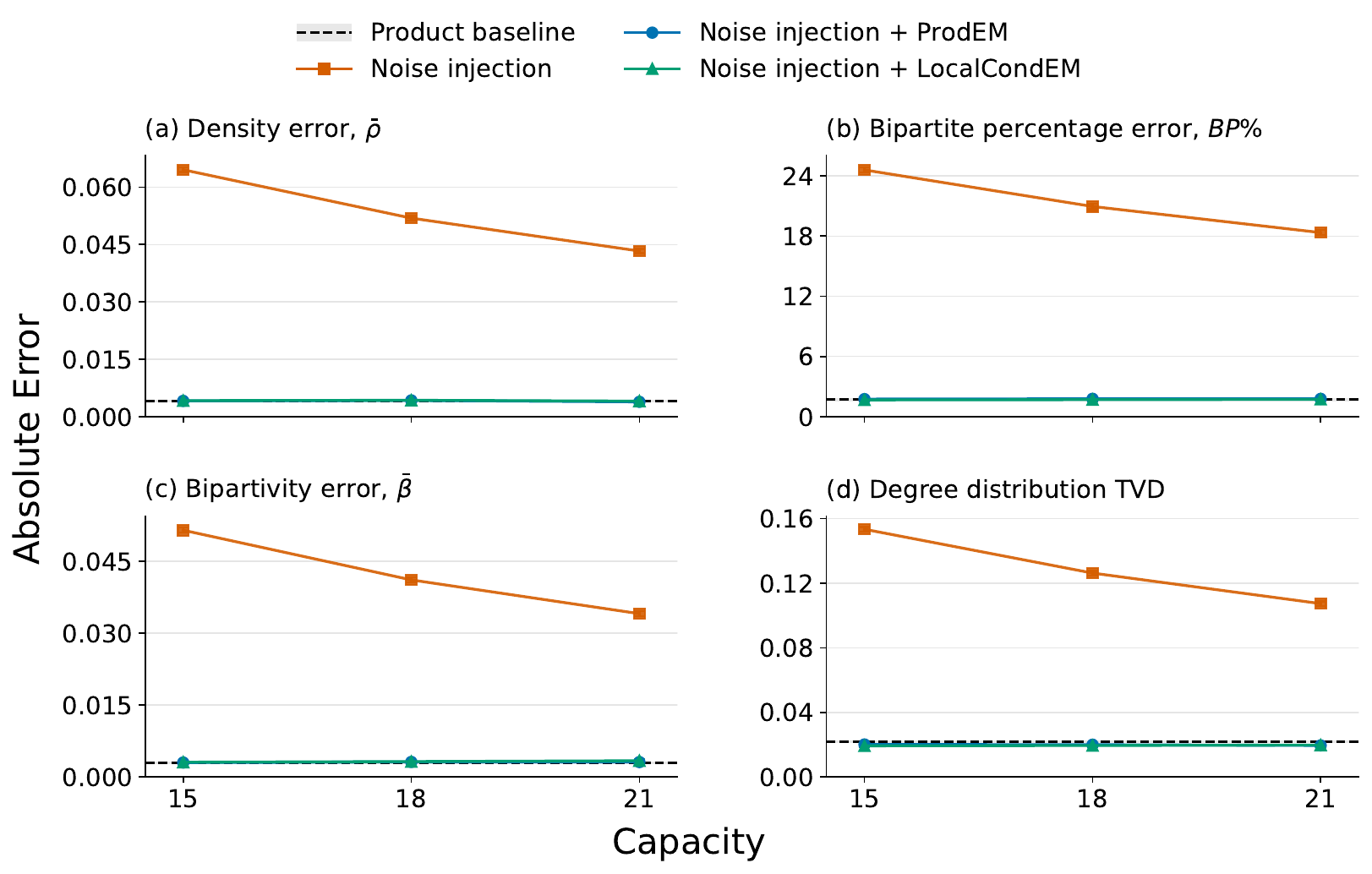}
    \caption{Sample quality metrics. Sample-quality errors averaged over six circuits and 50 repetitions per circuit. Error bars and the product baseline band show 95\% percentile bootstrap intervals for the mean, obtained from 5,000 resamples stratified by circuit. The dashed line denotes the product baseline mean. The deployment protocol without error mitigation has substantially larger errors across all metrics, although the error decreases with capacity. Both error mitigation methods reduce the errors close to the product baseline scale.}
    \label{fig:sample_quality}
\end{figure}

Figure \ref{fig:sample_quality} compares sample quality metrics as a function of the fragment capacity. The protocol without error mitigation exhibits large errors across all metrics, indicating that simply concatenating samples from capacity-limited fragments significantly distorts the generated graph distribution. Increasing the capacity reduces these errors but does not bring the protocol without error mitigation close to the product baseline. In contrast, both error mitigation methods reduce the errors to the product baseline scale across all capacities. Thus, error mitigation is necessary to recover sample quality in the tested settings, while increasing fragment capacity alone is insufficient.

\section{Random Pauli Z word expectation value analysis}\label{secAppendixRandomPaulis}

The samples from Section~\ref{sec4_1} are used to calculate random Pauli Z word expectation values. For each circuit, we used all 28 weight-1 and 378 weight-2 supports, and selected 500 distinct supports uniformly at random at each of weights 3 and 4. As in Appendix~\ref{secAppendix28QDetails}, ideal expectation values were estimated using 10,000 \texttt{Iqpopt} samples and held fixed during the bootstrap.

\begin{figure}
    \centering
    \includegraphics[width=0.85\linewidth]{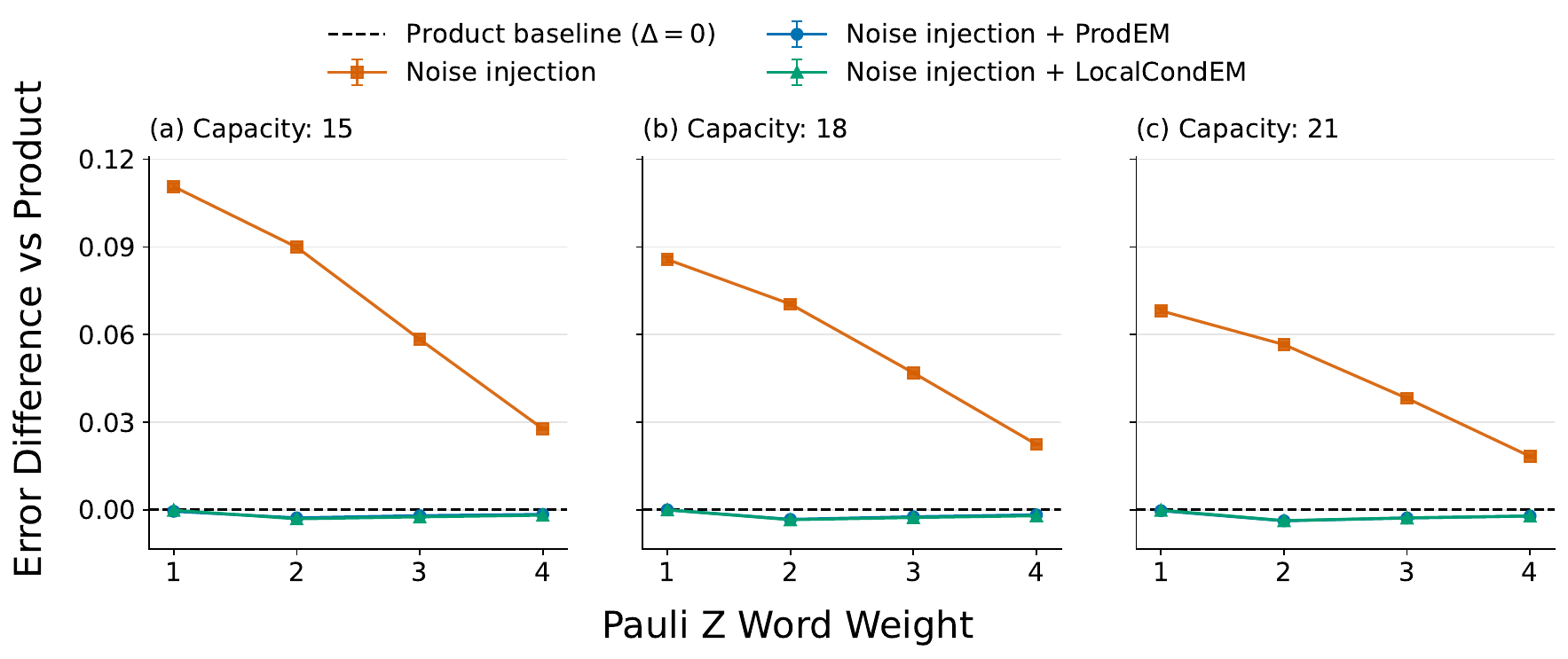}
    \caption{Random Pauli Z word expectation value errors relative to the product baseline. We plot $\Delta$=Err(method)-Err(Product) for uniformly random Pauli Z words of weights 1-4, aggregated over six circuits and 50 runs per circuit. Error bars show 95\% percentile bootstrap intervals from 5,000 replicates, independently resampling method and Product repetitions within each circuit. Negative values indicate lower error than the product baseline. The protocol without error mitigation is significantly worse than the baseline, while error mitigation recovers weight 1 statistics and slightly improves random higher-weight correlators.}
    \label{fig:random_correlators}
\end{figure}

Figure~\ref{fig:random_correlators} compares mean absolute errors for uniformly random Pauli Z word expectations of weights 1–4 relative to the product baseline, $\Delta=\text{Err}(X_{method})-\text{Err}(X_{product})$. The deployment protocol without error mitigation has significantly larger errors than the product baseline across all capacities, although the gap decreases with capacity and Pauli word weight. Both error mitigation methods substantially reduce this error, achieving errors close to those of the product baseline on weight 1 observables and slightly improving over the baseline for weights 2–4.

\section{IBM Heron experiment details}\label{secAppendixIBMDetails}

We provide details on hardware noise modeling, experimental resources, and performance metric calculation of the IBM Heron experiment. 

\subsection{Noise modeling}

\begin{figure*}
\centering
\begin{tabular}{c c}
  \includegraphics[width=0.35\textwidth]{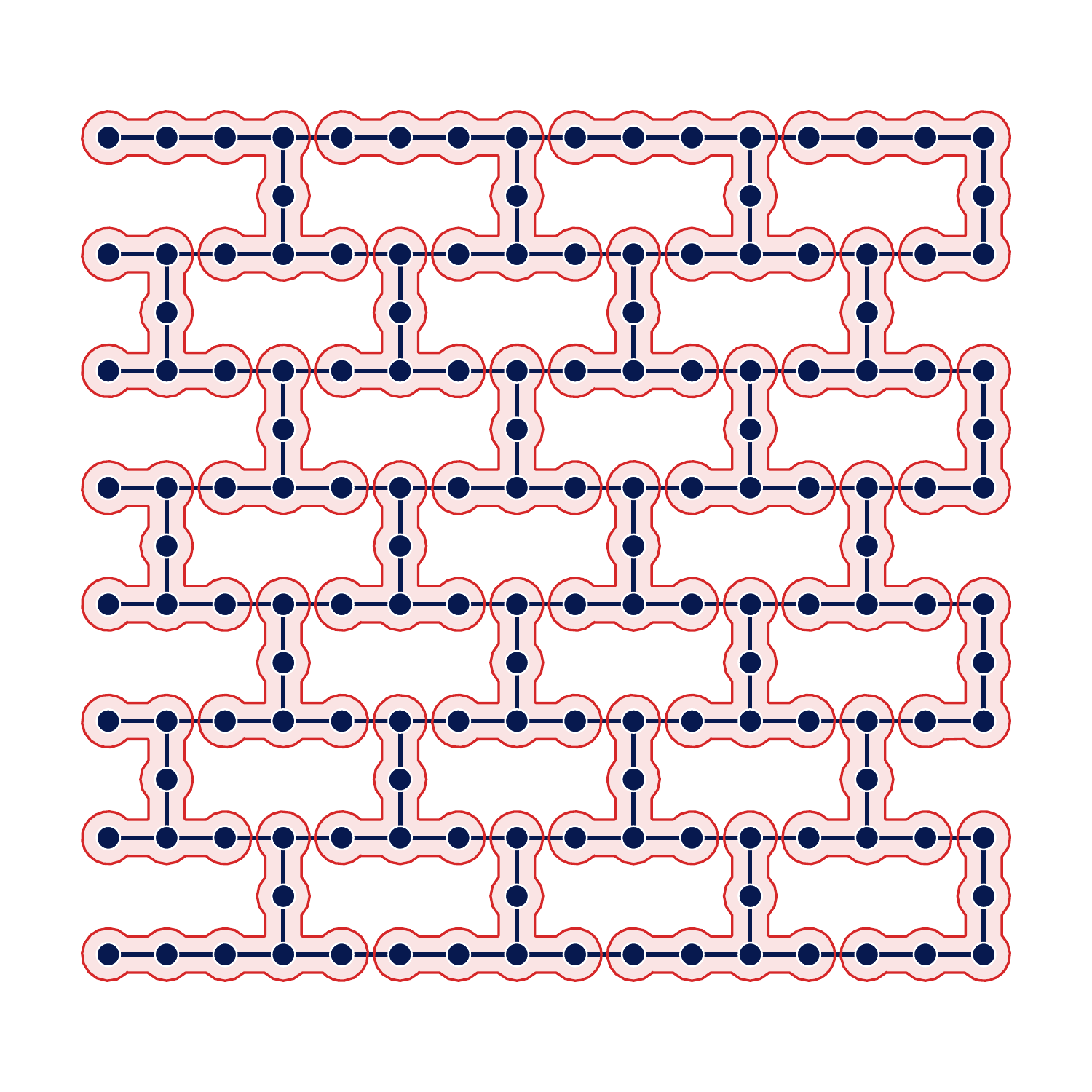} &
  \includegraphics[width=0.35\textwidth]{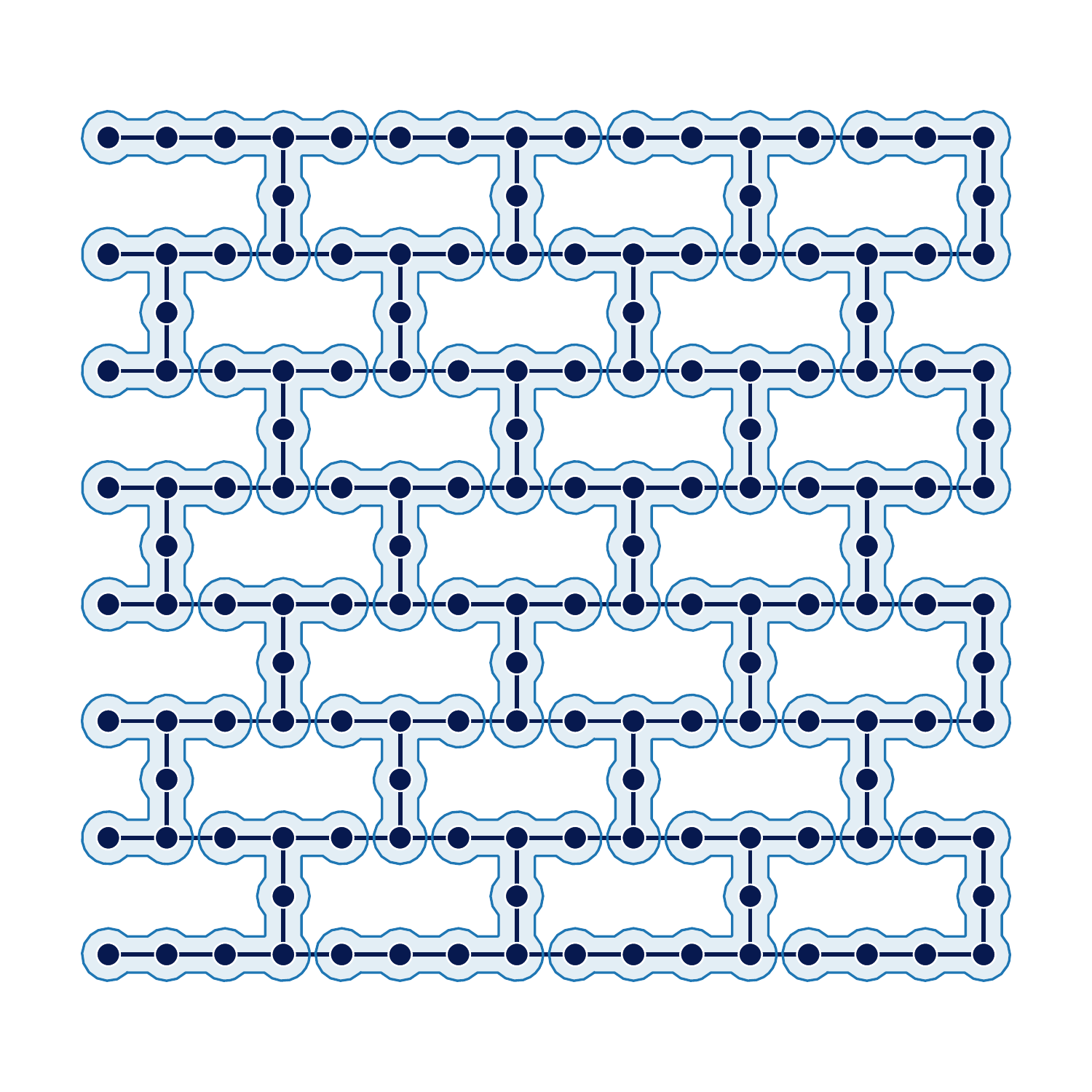}\\
  (a) & (b)
\end{tabular}
\caption{Partitioning \texttt{ibm\_marrakesh}'s qubit connectivity graph for noise modeling. (a) ``Regular" partition, (b) ``Flipped" partition}
\label{fig:partitions} 
\end{figure*}

The IQP circuits on each fragment consist of $RZ$ gates on each qubit with a random rotation angle sampled from $U[0, 2\pi]$ and $RZZ$ gates on every qubit pair connected by the hardware with a random rotation angle sampled from $U[0, \pi/2]$. The latter is chosen to make use of the fractional gates feature on IBM Heron processors, which allows $RZZ$ gates to be used as basis gates and reduces runtime \cite{ibmfractional}. 5 different sets of angle values are sampled for each partition, resulting in 10 different quantum circuits run on the hardware for noise modeling. Six are used for training, while the rest are used for validation. 

Each small circuit fragment is simulated using the Python library \texttt{PennyLane}, and the single-qubit Pauli noise channel is modeled in 3 different ways. (1) A single-parameter model, $\mathcal{N}_{QC}(\cdot) = (1-p)(\cdot)+pI/2$, (2) a two-parameter model, $\mathcal{N}_{QC}(\cdot)=\mathcal{N}_{p_x,0,0}\circ\mathcal{N}_{0,0,p_z}(\cdot)$, and (3) a three-parameter model, $\mathcal{N}_{QC}(\cdot) = \mathcal{N}_{p_x, p_y, p_z}(\cdot)$. The average $\text{MMD}^2$ across all different fragments of six training circuits between the simulated measurement distribution and the empirical distribution from the IBM processor is minimized by changing the noise parameters. The maximum mean discrepancy when comparing $n$-bit samples $x, y$ is calculated with the Gaussian kernel $k(x,y) = \frac{1}{|\Sigma|}\sum_{\sigma_i\in \Sigma}\exp\big({\frac{-|x-y|^2}{2\sigma_i^2}}\big)$, where $\Sigma = \{1/\sqrt{2}, n/4\sqrt{2}, n/2\sqrt{2}\}$. The single-parameter model is optimized through grid searching over 100 equally spaced points in [0, 0.05] and the two-parameter model is optimized by searching through the square grid where each parameter takes 50 equally spaced values in [0, 0.05]. The three-parameter model is optimized through Bayesian optimization using Gaussian processes, with 5 initial points and the negative expected improvement acquisition function being evaluated 50 times in total.

The minimum training objective and the validation objectives computed on four held-out circuits are similar across all noise models considered as detailed in Table \ref{tab:noise_fitting}. The resulting $\text{MMD}^2$ values are similar across the considered models. Thus, the single-parameter depolarization noise model is used in the IBM experiment.

\begin{table}%[<float-position>]
\caption{Noise model fitting results. The empirical measurement distribution from the hardware and the classically simulated Born distributions with the depolarizing/bitflip+dephasing/arbitrary 1-qubit Pauli noise channel model are compared using the maximum mean discrepancy. The training and validation mean $\text{MMD}^2$ values are taken over six and four circuits respectively. Each training circuit has different rotation angles and follows either the Regular or Flipped partition. Since the training and validation $\text{MMD}^2$ values are similar across all models, the depolarizing model is used for the noise injection-based deployment.}\label{tab:noise_fitting}
\begin{tabular}{c|c|c|c}
\toprule & Depolarizing & \begin{tabular}[x]{@{}c@{}}Bitflip\\+ Dephasing \end{tabular} & Arbitrary 1Q Pauli \\ 
\midrule Training Mean MMD$^2$ & 0.0110 & 0.0110 & 0.0110\\
Validation Mean MMD$^2$ & 0.0113 & 0.0114 & 0.0114 \\
Fitted Parameters & $p = 0.0404$ & \begin{tabular}[x]{@{}c@{}}$(p_x,p_z) =$\\$(0,0.0204)$ \end{tabular} & \begin{tabular}[x]{@{}c@{}}$(p_x,p_y,p_z) =$\\$(0.0100,0.0210,0.0002)$ \end{tabular}\\
\bottomrule 
\end{tabular} 
\end{table}

\subsection{Noise injection-based deployment}

A brief summary of resources is provided in Table \ref{tab:ibm_exp_resources}. During the noise injection-based deployment protocol, the initial injected noise strength was estimated using Equation~\ref{eq:usol}, which showed a success probability higher than 0.9 and was not fine-tuned. In fact, percolation caused 1015 out of 1024 tries to have clusters less than or equal to 156 qubits. Note that this occurs during classical preprocessing and does not result in any quantum resource overhead. Fragments of size 8 or less were classically simulated with the modeled noise. Fragments of size between 9 and 79 were paired by sorting by size then pairing the largest with the smallest. This was to prevent situations where extra SWAP gates are introduced during the transpilation process or when the pairing exceeds hardware capacity. The paired-up circuits were then run on the quantum computer at the same time. Larger fragments were run on their own. In total, 3362 circuit evaluations were performed on the quantum computer to obtain 1015 samples. The same samples from the noise injection-based deployment were used for error mitigation.

\begin{table}[]
\caption{IBM hardware experiment resource accounting. Out of 1024 attempts to run the protocol, 9 resulted in fragments larger than the available hardware. Fragments of size 8 or smaller were classically simulated, while fragments between 9 and 79 qubits were paired up and run in parallel. This resulted in 3,362 hardware calls to obtain 1015 samples.}
    \centering
    \begin{tabular*}{\textwidth}{@{\extracolsep{\fill}}lc}
    \toprule Quantity & Value \\
    \midrule 
    Empirical $p_p$ & 0.2424\\
    Target IQP Qubits & 316\\
    Hardware Qubits & 156\\
    Target/Hardware Ratio & 2.03\\
    Deployment Attempts & 1024\\
    Capacity Discarded Attempts & 9\\
    Small Fragments of Size $\le 8$ & 28,794\\
    Medium Fragments of Size $\in [9, 79]$ & 6284\\
    Large Fragments of Size $\ge 80$ & 220\\
    Number of Executed Hardware Circuits & 3362\\ \bottomrule
    %\end{tabular}
    \end{tabular*}
    \label{tab:ibm_exp_resources}
\end{table}

For the correlator comparisons, all edge pairs, distance-two pairs, nonlocal pairs, connected triples, and connected 4-tuples were evaluated unless their number exceeded 1,000, in which case 1,000 were selected at random. Nonlocal pairs were defined as pairs at graph distance at least four. Ideal expectation values were estimated using \texttt{Iqpopt} with 10,000 samples and held fixed throughout the analysis. For each observable group, we calculated the observed mean absolute error difference between each method and Product. Complete bit-strings were independently resampled with replacement within the two sample sets, retaining their original sizes, for 5,000 bootstrap replicates. We subtracted the bootstrap mean from the replicate differences and added the resulting 2.5th and 97.5th percentile deviations to the observed difference to obtain recentered bootstrap intervals. These intervals condition on the estimated ideal correlators and selected observables, excluding reference-estimation uncertainty, hardware drift, and calibration variability.

For sampling method testing, $\text{MMD}^2$ was calculated using \texttt{Iqpopt} where 5,000 Pauli Z operators were sampled per kernel bandwidth and 5,000 Monte Carlo samples were used for each expectation value evaluation. The kernel used for $\text{MMD}^2$ followed the setting used for the noise modeling. 500 bootstrap datasets were created by resampling bit-strings with replacement, then the $\text{MMD}^2$ was calculated 5 times using the same bootstrap dataset. These 5 values were averaged, then the 500 averages were used to calculate the 95\% confidence intervals.

\end{appendices}

%\section*{References}
\bibliographystyle{iopart-num}
\bibliography{bibliography}

\end{document}